# Strategies educators can use to counter misinformation related to the quantum information revolution

Jaya Shivangani Kashyap and Chandralekha Singh
*Department of Physics and Astronomy, University of Pittsburgh, 3941 O'Hara St, Pittsburgh, PA, 15260*

**Abstract:**

Remarkable advances in quantum information science and technology (QIST) have taken place in recent years. However, they have also been accompanied by widespread misinformation. This paper provides suggestions for how educators can help students at all levels and especially early learners including those at the pre-college and college levels learn key QIST concepts so that they are less likely to be misinformed, e.g., by online unvetted resources. We discuss findings from interviews with five college educators, who are quantum researchers, about their views on countering misinformation in QIST and provide suggestions for how educators can help their students learn QIST concepts so that they do not become misinformed.

## Introduction

Over the past few decades, the second quantum revolution—often referred to as the quantum information revolution—has progressed at an extraordinary pace. This progress has been driven by our ability to precisely control and coherently manipulate quantum states of matter, leveraging principles such as quantum superposition and entanglement. Quantum technologies hold the potential to revolutionize computing, communication, and sensing capabilities. The rapid advancement of quantum information science and technology (QIST), particularly accelerated by the European Commission Flagship kick off and the passage of the US National Quantum Initiative Act, presents exciting opportunities to prepare students to become leaders in this transformative field [1-3].

Like many high-profile subjects, popular treatments of QIST are filled with misinformation in media and the broader community, especially surrounding key concepts including entanglement, teleportation, and quantum computers. For example, there is misinformation that quantum computers will become cost effective and completely replace classical computers that we use routinely. A related misinformation is that quantum computers will provide exponential speed up for *all* problems that a classical computer can solve. Only certain types of problems have quantum advantage and can be solved faster on a quantum computer and even fewer problems will have an exponential speedup. For example, quantum computers will provide exponential speedup for factoring products of large prime numbers that is at the heart of the encryption scheme that keeps our credit card transactions safe on the internet [4]. Another related misinformation is that quantum computers will imminently break the current encryption schemes that make our credit card transactions safe, when we are not likely to achieve such milestones at the very least until several decades into the future (also, there are already post-quantum encryption schemes that do not require factorization of products of large prime numbers that can be used to keep encryption safe even if large fault-tolerant quantum computers were built). There is also misinformation that if we build large quantum computers, they will be able to solve all of the world's problems. Although the problems that quantum computers can solve are important, quantum speedup only applies to a small subset of the set of all problems that can be addressed with computers.

We conducted thirteen 1-1.5 hour individual interviews with college educators related to different aspects of QIST, and most of them provided their views on how to promote diversity in this field [5] and reflected on other critical issues in QIST education such as how to ensure that students are not misinformed, which is discussed here. To get the salient points across and minimize repetition, here we focus on only five educators' views that capture the essence of persistent misinformation related to QIST concepts and how to ensure that students are not misinformed. Thus, this paper focuses on reflections and suggestions from those five college QIST educators about what can be done to counter misinformation and misconceptions related to the QIST concepts. These five interviewed educators were from different parts of the world, i.e., one was from UK, another from Australia and three were from the US. Since the educators had similar suggestions for countering misinformation, we do not identify which educator was from which country. Two of the educators had taught courses related to QIST at both the undergraduate and post-graduate levels whereas three of them had taught undergraduate level courses. They are also actively involved in informal outreach and education for early learners. However, none of them conducts physics education research. These educators discussed misinformation, e.g., about quantum entanglement implying that we can do faster than light communication, quantum teleportation being like Star Trek (material instead of quantum information being teleported from one place to another), and quantum computers being several classical computers running in parallel [6]. These interviewed educators are passionate about public/early understanding of QIST concepts. They were interviewed using a semi-structured think-aloud protocol about various issues related to QIST education including those pertaining to what educators can do to counter misinformation about the quantum information revolution discussed here.

Below, we discuss some excerpts that exemplify QIST researchers/educators' reflections and suggestions for countering misinformation/misconceptions related to QIST.

## Views of interviewed educators

Educators interviewed highlighted the need to carefully address aspects of QIST concepts that might be confusing or could contribute to misinformation. Educator A reflected on this issue in the context of entanglement implying faster than light communication emanating from outcomes of measurements of spatially separated entangled entities being correlated saying, "We can't send information faster than light…We can't use entanglement [to do that]". To counter misinformation, he suggested that "one has to emphasize that whatever else you may think about how the universe works, it doesn't allow us to send signals faster than light".

Regarding misinformation related to teleportation he said, "maybe part of the problem is the word teleportation because it suggests material moving. I defend that word though. I think it's a correct word. But what do you do about it?...just to emphasize…teleportation is to move a *quantum state* from one place to another". He further added, "When people say, oh, if you're just moving information around, what's the big deal? But it is a big deal because of the *kind* of information it is, and that's because of the delicacy of it that…you can't just measure it, and you can't copy it. So it's really crucial to emphasize that the *kind* of information that's being conveyed - it's not what you're used to. It's not what a fax machine can do…if you teleport a quantum state from one place to another, that state can't still be at the original place…you can't teleport a quantum state from A to B and…the quantum state still be at A the way you can with a fax machine".

Educator B, who acknowledged that it can be annoying when you see misinformation, e.g., about teleportation being just like you see it on Star Trek, said, "…[as] academics in the area, we have a good opportunity, in a sense, to help correct some of that, to help sort of publicly provide, you know, assistance that helps to manage the expectations, and…explain why we're really excited about what these things can do."

Educator C, who has made videos on YouTube explaining QIST concepts, e.g., teleportation, for the public noted, "[when] people see them…at least they start to hear those basic things. This is not faster than the speed of light. You understand how you do quantum key distribution where you teleport a "bit" of information, but you need the classical bit to go along with that, and of course the classical bit goes at the speed of light, not faster than that…"

Educator D started with what she does in her own college classes at various levels saying, "So I make sure to point out which parts are misinformation. And sometimes I even share that humorous cartoon by Scott Aaronson [and Zach Weinersmith] [7]. Have you seen that? *If you don't talk to your kids about quantum computing, someone else will*…this is hilarious. [In the comic clip], the parent and the child are trying to talk about quantum information. And they start with first, kind of the basic principles…those things that you would hear in newspapers. And they're like, well, yeah, but not really. And then they go deeper and deeper. And you see how many misconceptions are [discussed in those]…for example, whether all calculations are happening in parallel [on a quantum computer]."

She noted that she loves to answer her students' questions in office hours if they are troubled about the same type of misinformation that is out there on the internet or are generally confused. She confessed that when she is interviewed for an article by the media, she worries greatly because the interviewer is likely to not get it right and frame it incorrectly propagating misinformation.

Educator C further noted that to counter misinformation, "Take every opportunity you have, especially with non-standard academic ways of disseminating your knowledge, to talk to the people. One thing I found very, very useful and important over the years is [that] rather than trying to do it [disseminating] yourself, try to hook on to people who do it professionally." For example, he emphasized the importance of quantum educators partnering with professional youtubers saying, "that is very efficient…if you can insert your message into that pre-established large following, that's how you reach the largest number of people at the minimum amount of effort on your part." He stated that he has developed and disseminated many QIST videos that have been made in collaboration with youtubers who want to monetize the public interest surrounding QIST.

Educator E, who regularly teaches college quantum courses for non-STEM majors, emphasized the importance of reaching diverse students with vetted resources that are well thought out and carefully developed to counter misinformation saying, "a lot of scientists post videos on YouTube, you probably know Sabine Hossenfelder [8]…she puts out a lot of really educational…videos that are accessible…[includes topics] like time travel wormholes, quantum computing, Bell's inequalities, determinism, all kind of interesting things…she's probably the best person out there who tries to talk the truth, tells the truth". He also noted that QIST educators and experts participating actively in "online media" and "other science websites" can play a role in debunking misinformation saying, "you can also write…editorials to your local newspaper, and I've done that a few times trying to debunk certain headlines that are out there".

Educator E also emphasized the importance of getting younger students interested in science so that they would like to delve deeper into the QIST concepts beyond the flashy headlines which can sometimes be deliberately misleading to increase views saying, "but maybe the most important thing is to just get younger people interested in science, so they're willing to read stories about it, and then they can find the ones that are telling the true story, not just the headlines".

Educator B also emphasized that "it's important to try to do sort of public communication going to…science centers or [giving] general public talks, public lectures… [so] people can come along to ask those questions". He felt that this can provide opportunities to clarify, e.g., that if they saw some result, it doesn't mean faster than light communication, while still clarifying that QIST concepts are still really interesting.

## Discussion and Summary

Since QIST is a rapidly growing interdisciplinary field with major implications for the future workforce, educators teaching introductory college courses should consider incorporating foundational QIST concepts in their courses as appropriate. Here we discussed reflections of five college educators deeply involved in QIST research about what educators can do to counter misinformation related to QIST concepts.

Their suggestions for helping students learn QIST concepts and counter misinformation can be summarized in the following categories:

- Educators reflecting deeply about the challenging aspects of QIST concepts, e.g., entanglement, teleportation, quantum vs. classical computers, etc., that lead to misinformation and
  - communicating about the concepts in a manner that minimizes incorrect interpretation
  - developing curricula, pedagogies to improve understanding at all levels
- Educators sharing with their students at all levels vetted resources on the internet, e.g., videos, simulations, articles and other tools, so that they can learn from them as self-paced learning tools or as part of a course in which the educators and students will also discuss those ideas in class
- Educators introducing QIST concepts in courses for non-science majors
- Educators engaging with the public by participating in outreach to science centers and giving public lectures
- Educators engaging with media, writing articles for online forums, blogs and developing videos

Deep reflections about how to teach these QIST concepts effectively on the part of educators is critical. For example, educators should highlight that quantum teleportation inherently adheres to the speed of light limit, as its process relies on both a quantum entangled channel and a classical communication channel. Moreover, it is important to understand and connect with the prior knowledge of early learners, but recognize that over-simplification can lead to incorrect mental models [6]. For example, if quantum teleportation is taught with a demonstration in which a person was initially in one place and was transported to another place, it can make students develop a mental model that quantum teleportation is about material (instead of quantum information) transport from one place to another (although certain aspects of this model such as quantum information vanishing from the original place are correct).

Engaging and motivating approaches to introducing foundational concepts relevant for the quantum information revolution to introductory students using a two-state system can be valuable. For example, educators can give students tools to visualize QIST concepts [9-11] as well as help them reflect upon different physical realizations of qubits (e.g., polarization states of photons, spin-1/2 system, electronic degrees of freedom of atoms) and their applications [12]. Educators can also include hands-on activities involving polarization of light with students doing investigations with three polarizers (with different orientations of polarization axis) and then having a class discussion about how the classical beam of light differs from single photons. Furthermore, giving introductory students opportunities for kinesthetic exploration and game-based approaches to learn basic QIST concepts can make learning fun and meaningful. For example, Hahn and Gire [13] devised approaches to guide students to learn basic quantum concepts by waving their arms. Lopez-Incera et al. [14, 15] developed several engaging games using qubits in which students play the role of particles and scientists simulating a real lab to help them learn QIST concepts. These games give students opportunities to collect and analyze data and understand how these observations are different from what they would observe in a classical situation. These games make it easier for students to have a good mental model of quantum concepts in a motivating and engaging environment. [14, 15]. Lopez-Incera et al.'s games give students the opportunity to understand the probabilistic nature of quantum measurement as well as quantum entanglement [14]. They have also used this approach to help students learn about decoherence, Bell's inequality and quantum cryptography and found that it helps students understand the significance of quantum concepts and how they are different from classical concepts [14, 15]. Marckwordt et al. built on Lopez-Incera et al.'s game-based approaches and developed a dodgeball-based game to teach quantum entanglement [16]. Other approaches to teaching two-state systems, e.g., comparing and contrasting quantum and classical concepts [17, 18] and including diagrammatic approaches such as that by Rudolph with basic mathematics can be valuable [19].

Furthermore, specifically focused on secondary school, Michelini et al. put forward a comprehensive proposal for quantum physics [20]. Bitzenbauer conducted an investigation to get practitioners' views on new teaching material for introducing quantum optics in secondary schools [21] and Toth et al. developed and implemented a phenomenological teaching-learning sequence that improved lower secondary school students' views of polarization of light [22]. Montagnani et al. developed an experiential program on the foundations of quantum mechanics for final-year high school students [23] and Hennig et al. developed a teaching-learning sequence to promote secondary school students' learning of quantum physics using Dirac notation [24]. Magdalena et al. mapped out future directions in modern physics education which includes quantum concepts [25]. Rodriguez et al. emphasized designing inquiry-based learning environments for teaching quantum in secondary schools [26]. Stadermann et al. developed easy-to-implement teaching materials that link quantum physics to the nature of science [27]. Schalkers et al. developed a pedagogical model to explain Grover's algorithm with a colony of ants for making quantum technology accessible [28].

To get students to think critically about QIST concepts using some of these innovative approaches so that students are not misinformed by incorrect concepts, e.g., through online unvetted resources, after introducing their students to these concepts, educators can assign students some homework and classroom activities in which students must read (watch) some articles (videos) on the web with misleading headlines related to those concepts. Along with misleading articles, students can be assigned some articles/videos that help elucidate those concepts well for introductory learners. Then there can be class discussions comparing the content of both types of articles and why it is important to go beyond the flashy headlines to understand these fascinating QIST concepts. Students can also be given more autonomy and opportunities to search for articles/videos on the web that are propagating misinformation and those that are teaching the same QIST concepts well. Then they can be asked to present their findings comparing different sources they investigated in the class to other students and teachers. Students can be assigned these activities in small groups so that they can brainstorm among themselves before presenting and discussing their findings with the entire class. Educators can provide appropriate support to their students as needed to reinforce QIST concepts and help them learn to differentiate between the articles/videos propagating misinformation and those helping them to learn those concepts well.

Finally, education researchers should investigate the different origins of misinformation about various QIST concepts and continue to develop curricula and pedagogies as well as evaluate their effectiveness to help learners at different levels with different prior preparations learn these exciting but challenging concepts so that they are not misinformed.

## Acknowledgement

This research is supported by the US National Science Foundation Award PHY2309260. We thank all faculties who participated.

## Ethical Statement

This research was carried out in accordance with the principles outlined in University of Pittsburgh ethical policy, the Declaration of Helsinki, and local statutory requirements. The educators provided consent for use of the interview data for research and publication, and consent for quotes to be used.

**About authors**

Jaya Shivangani Kashyap is a Ph.D. student pursuing research in physics education in the Department of Physics and Astronomy at the University of Pittsburgh.

Chandralekha Singh is a Distinguished Professor of Physics in the Department of Physics and Astronomy at the University of Pittsburgh.